\documentclass[a4paper]{jpconf}
\usepackage{graphicx}
\usepackage{amsmath,amsthm, amssymb, latexsym}
\usepackage{cite}
\usepackage{color}

\begin{document}
\title{Approaching the QCD phase diagram for $N_f=2+1$ and $N_f=2+1+1$ quark flavors}

\author{Christian A Welzbacher$^1$, Christian S Fischer$^1$ and Jan Luecker$^{2,3}$}

\address{$^1$Institut fuer Theoretische Physik, Justus-Liebig-Universitaet Giessen, Heinrich-Buff-Ring 16, D-35392 Giessen, Germany.}
\address{$^2$Institut fuer Theoretische Physik, Universitaet Heidelberg, Philosophenweg 16, D-69120 Heidelberg, Germany.}
\address{$^3$Institut fuer Theoretische Physik, Goethe-Universitaet Frankfurt, Max-von-Laue-Strasse 1, D-60438 Frankfurt/Main, Germany}

\ead{Christian.A.Welzbacher@theo.physik.uni-giessen.de}

\begin{abstract}
Recent results for the QCD phase structure at finite temperature and light-quark chemical potential are summarized, where 
 the cases of $N_f = 2 + 1$ and $N_f = 2+1+1$ dynamical quark flavors are considered. Order parameters for the 
 chiral and deconfinement transitions are obtained from solutions of a coupled set of truncated Dyson-Schwinger equations 
 for the quark and gluon propagators of Landau gauge QCD. Based on very good agreement with lattice QCD for zero chemical potential 
 the phase diagram in the whole T-$\mu$ plane and the appearance of a putative critical end-point at large chemical potential are studied. 
\end{abstract}

\section{Introduction}

In recent years, experimental and theoretical interest in heavy ion collisions and the phase diagram of QCD grew rapidly.
On the experimental side facilities such as RHIC at BNL, ALICE at LHC and CBM at the future FAIR facility (will) search for possible signals of phase transitions.
On the theoretical side lattice QCD has established a crossover transition at small chemical potential, see e.g. \cite{Borsanyi:2010bp,Bazavov:2011nk} for $N_f=2+1$ flavors. In our work we 
approach the QCD phase diagram with the functional method of Dyson-Schwinger equations and put particular emphasis on the possible influence of the charm quark \cite{Fischer:2014ata}.
The contribution of the charm quark to the equation of state starts to become significant at top LHC energies. However, possibly even at smaller temperatures 
close or above the light-quark crossover region, effects of charm quarks, although predicted to be small by perturbation theory \cite{Laine:2006cp}, may not be entirely negligible. 
Within lattice QCD preliminary results on calculations with $N_f=2+1+1$ flavors for transition temperatures and the equation of state using staggered \cite{Bazavov:2013pra,Borsanyi:2011} and Wilson type quarks \cite{Burger:2013hia} are available.
Here we summarize recent results for the phase diagram at zero and finite chemical potential within the framework
of Dyson-Schwinger equations. At zero chemical potential we compare with lattice results for the quark condensate
and the unquenched gluon propagator and discuss our results for the critical end point for $N_f = 2 + 1$ and 
$N_f = 2 + 1 + 1$ quark flavors. 

\section{Framework}
\begin{figure}[h]
\begin{minipage}{20pc}
\includegraphics[width=16pc]{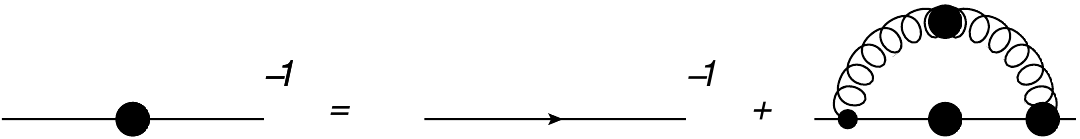}
\vspace{2.5pc}
\caption{\label{fig:qdse}The quark Dyson-Schwinger equation.}
\end{minipage}\hspace{2pc}%
\begin{minipage}{20pc}
\includegraphics[width=16pc]{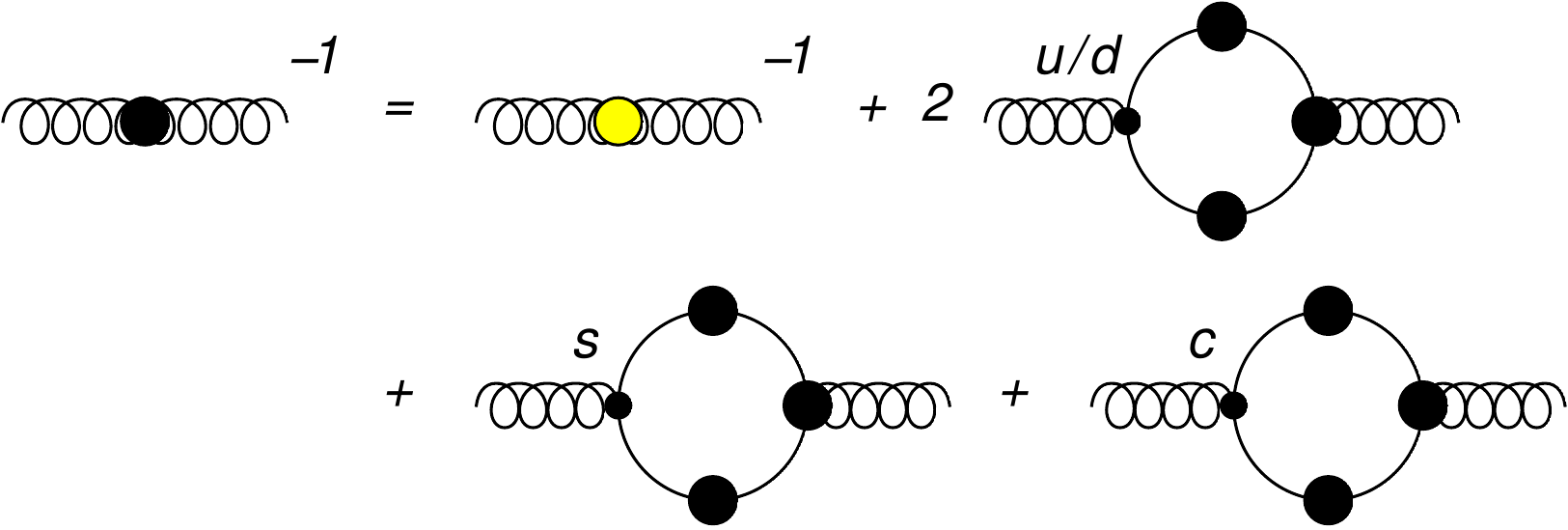}
\caption{\label{fig:gdse}The truncated gluon \newline Dyson-Schwinger equation.}
\end{minipage} 
\end{figure}

Dyson-Schwinger equations (\emph{DSEs}) are the equations of motion for the n-point functions of QCD. 
In Fig.~\ref{fig:qdse} we show the DSE for the quark propagator, which is the foundation of an 
infinite tower of coupled integral equations. These are in principle ab initio, if solved completely 
and self-consistently. In practice, however, in most cases one needs to impose a truncation using input 
from other sources such as symmetries, constraints or other methods like lattice QCD.
\newline
The truncation used in this work has been developed over the past few years (see \cite{Fischer:2009wc,Fischer:2010fx,Fischer:2012vc,Fischer:2013eca,Fischer:2014ata}) and consists of two key points. The first part 
of the truncation is depicted in Fig.~\ref{fig:gdse}, showing the approximation in the DSE for the 
Landau gauge gluon propagator. In order to calculate this quantity we use input from lattice QCD for 
the quenched gluon propagator (denoted by the yellow dot), replacing diagrams with internal gluons and 
Faddeev-Popov ghosts. To this we add a quark loop for each flavor.

This approach neglects all the second order unquenching effects in the Yang-Mills diagrams but ensures that we take the important temperature effects of the unquenched gluon into account. Additionally the unquenched gluon
propagator inherits a dependence on the temperature and the chemical potential via the quark loops, giving a contribution to the thermal mass.
The gluon propagators at finite temperature $T$ and quark-chemical potential $\mu$ is given by
\begin{equation}
\label{eq:gProp}
D_{\alpha\beta}(p) = P_{\alpha\beta}^L(p)\frac{Z^L(p)}{p^2} + P_{\alpha\beta}^T(p)\frac{Z^T(p)}{p^2}\,,
\end{equation}
with momentum $p=(\omega_n,\vec{p})$, $\omega_n= 2 n \pi T$ for bosons. The projectors $P_{\alpha\beta}^{L,T}$ are longitudinal (L) and transverse (T) with respect 
to the heat bath and given by
\begin{eqnarray}
P_{\alpha\beta}^T &=& \left(1-\delta_{\alpha 4}\right)\left(1-\delta_{\beta 4}\right)\left(\delta_{\alpha\beta}-\frac{p_\alpha p_\beta}{\vec{p}^{\,\,2}}\right), \label{eq:projT} \\
P_{\alpha\beta}^L &=& P_{\alpha\beta} - P_{\alpha\beta}^T \,, \label{eq:projL}
\end{eqnarray}
where $P_{\alpha\beta}$ is the transverse projector with respect to the four momentum.\newline
The second element of our truncation is an approximation for the full quark-gluon vertex. Information
on this vertex can be gathered from lattice QCD or from solutions to its DSE. Corresponding studies
at zero temperature and chemical potential are under way; for recent works see e.g. 
\cite{Williams:2014iea, Aguilar:2014lha} in the DSE framework and \cite{Mitter:2014} in the functional 
renormalization group approach. At finite temperature, however, information from these sources is not
yet available. We therefore use a functional form for this vertex which combines information
from the well-known perturbative behavior at large momenta and an approximate form of the Slavnov-Taylor 
identity at small momenta studied long ago by Ball and Chiu \cite{Ball:1980ay}.

The explicit form of this approximate expression for the vertex is discussed in Refs.\cite{Fischer:2010fx,Fischer:2012vc,Fischer:2014ata} and shall not be repeated here for brevity. It contains one open parameter, called $d_1$
in the following, which controls the infrared strength of the vertex and sets the temperature scale.
Furthermore, via the Ball-Chiu construction, it contains dressing functions of the two attached quarks.
Therefore the vertex is flavor, as well as temperature and chemical potential dependent. 
To investigate the influence of the charm quark, we developed two approaches to set the IR-strength of the quark-gluon vertex as well as the physical masses of the quarks. The first one (called Sets A$_{Nf}$) connects
our calculations to lattice QCD, since the IR-strength and light-quark mass is fixed to reproduce the (pseudo-) critical temperature $T_C \approx 155$ MeV from \cite{Borsanyi:2010bp} for $N_f$=2+1 flavors at zero chemical potential. Since in a crossover region there are different definitions 
of a critical temperature, we point out that the chosen one is defined via the inflection point of the chiral condensate. The mass of the strange quark is chosen to be 27 times the light-quark mass.
To this setup we merely added a charm quark with a mass of 300 MeV at a renormalization point of 80 GeV, without adjusting the IR-strength. In our second approach (called Sets B$_{Nf}$) we set the IR-strength in the quark-gluon vertex self-consistently by fixing the pion decay constant, as well as the pion, kaon and $\eta_C$ mass in the 
very same truncation via the Bethe-Salpether equation, separately for $N_f$=2+1 and $N_f$=2+1+1. This was possible due to recent progress in the Bethe-Salpether framework \cite{Heupel:2014ina}.

To describe the (pseudo-) critical temperature of our system, we calculate the chiral condensate 
\begin{equation}
    \langle\bar{\psi}\psi\rangle_f = Z_2 Z_m  N_c  T\sum_n\int\frac{d^3p}{(2\pi)^3}\mathrm{Tr}_D\left[S^f(p)\right],
    \label{eq:condensate}
\end{equation}
where $S^f(p)$ is the calculated quark propagator for flavor f, $Z_2$ and $Z_m$ are the corresponding wave-function and quark mass renormalization constants, respectively, and $N_c=3$ is the number of colors. Any quark flavor
with non-zero bare quark mass leads to a condensate which is quadratically divergent and needs to be regularized. To circumvent this divergence one considers the regularized chiral condensate
\begin{equation}
\Delta_{l,s} = \langle\bar\psi\psi\rangle_l - \frac{m_l}{m_s}\langle\bar\psi\psi\rangle_s\,.
\label{eq:cond_renorm}
\end{equation}
In our work we extract $T_C$ either by considering the maximum of the derivative of this quantity with respect to temperature (\emph{inflection point}) or with respect to the mass of the light quarks 
(\emph{chiral susceptibility}).

\section{Results}
\begin{figure}[h]
\begin{minipage}{20pc}
\includegraphics[width=18pc]{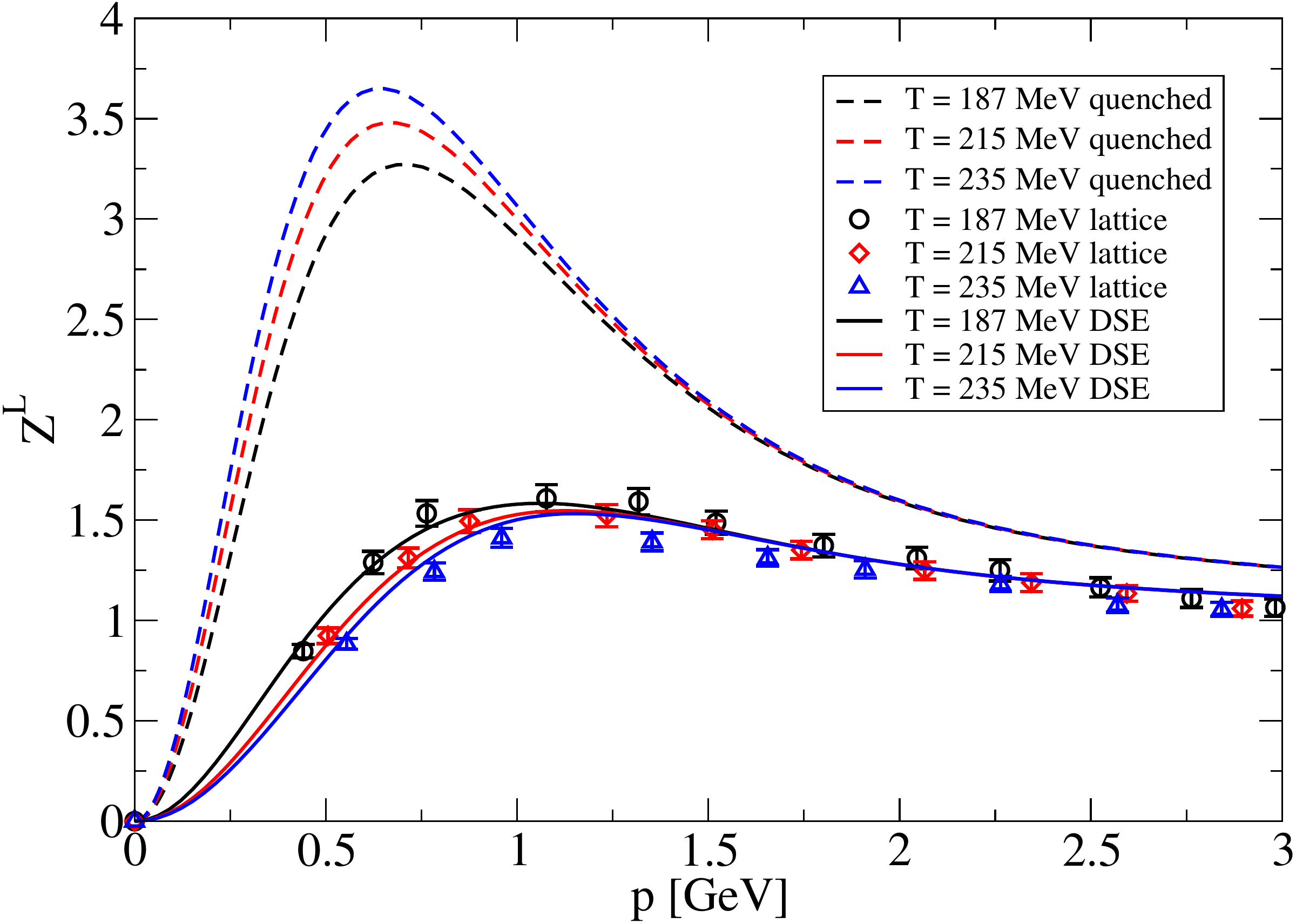}
\caption{\label{fig:ZlNf2}Electric part of the gluon \newline dressing functions for N$_f$=2, $m_\pi$=316 MeV.}
\end{minipage}\hspace{0pc}%
\begin{minipage}{20pc}
\includegraphics[width=18pc]{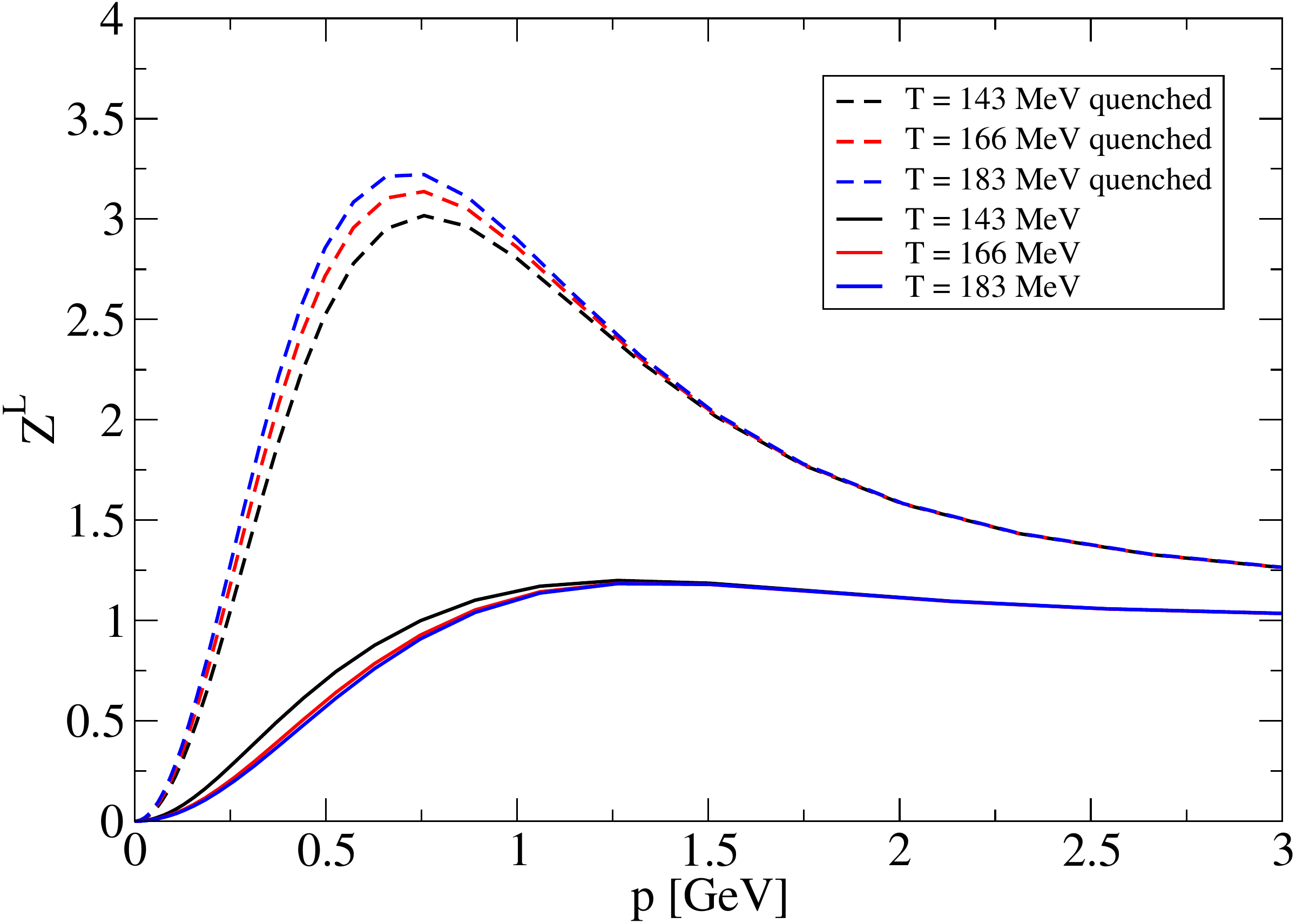}
\caption{\label{fig:ZlNf3}Electric part of the gluon \newline dressing functions for Set $A_{2+1}$.}
\end{minipage} 
\end{figure}

In this section we show the results calculated within the framework discussed in the previous section. In Fig.~\ref{fig:ZlNf2} we compare the electric part of the gluon propagator for N$_f$=2 and a pion mass of 316 MeV, fixed in the manner of Set B, 
with lattice data from \cite{Aou:2013}. Note that our result, determined first in \cite{Fischer:2012vc}, has
been a prediction, verified later by the lattice results. Therefore the quite good agreement for the unquenched 
gluon, featuring an inversion of the temperature ordering when going from the quenched to the unquenched case 
serves as a further justification for our truncation scheme. In Fig.~\ref{fig:ZlNf3} the same quantity is shown, in this case for Set A$_{2+1}$. We observe the same inversion of the temperature ordering as well as the suppression of the maximum after the unquenching.

\begin{figure}[h]
\begin{minipage}{20pc}
\includegraphics[width=18pc]{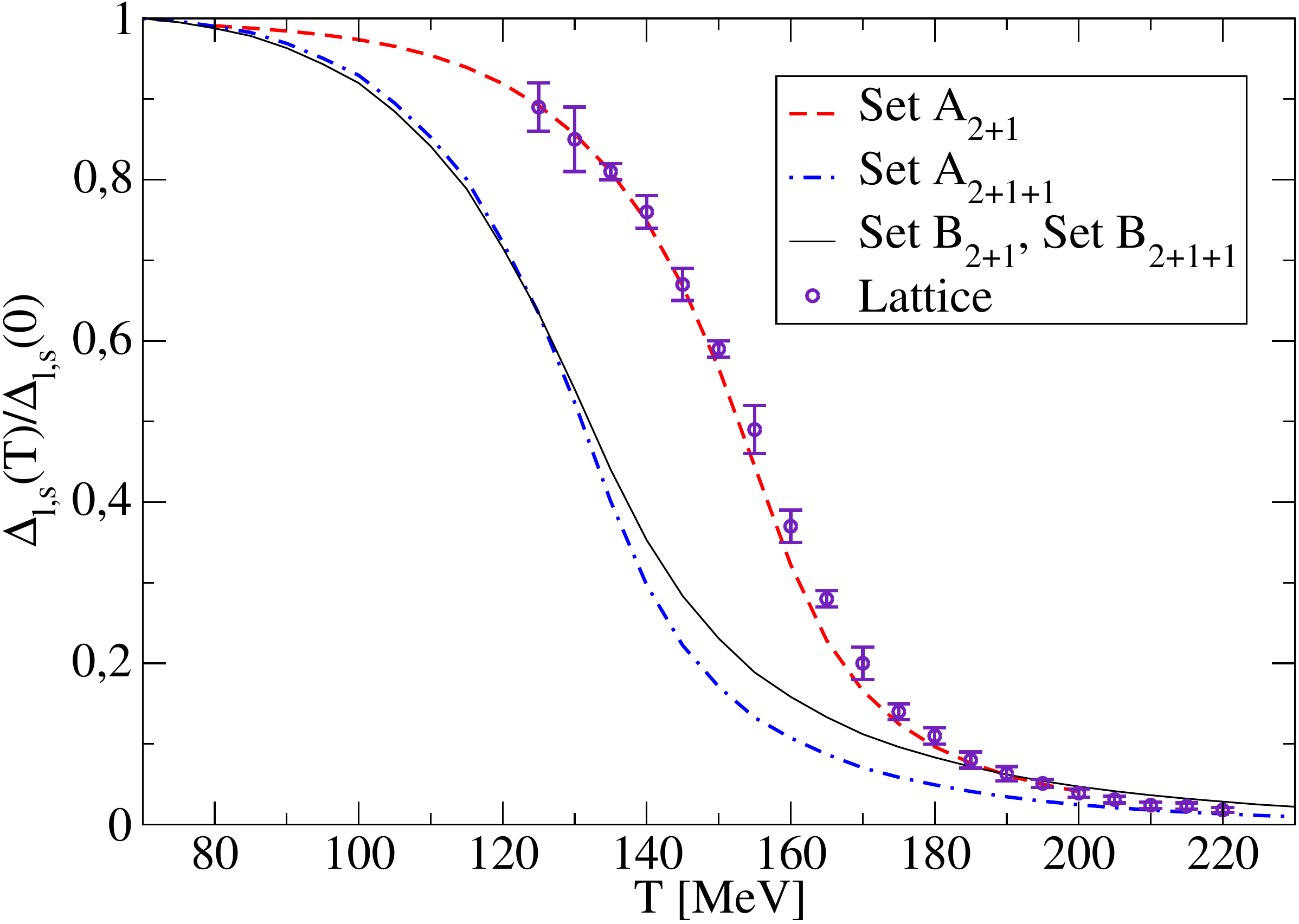}
\caption{\label{fig:mu000}$\Delta_{l,s}$ at $\mu$=0 MeV \cite{Fischer:2014ata}.}
\end{minipage}\hspace{0pc}%
\begin{minipage}{20pc}
\includegraphics[width=18pc]{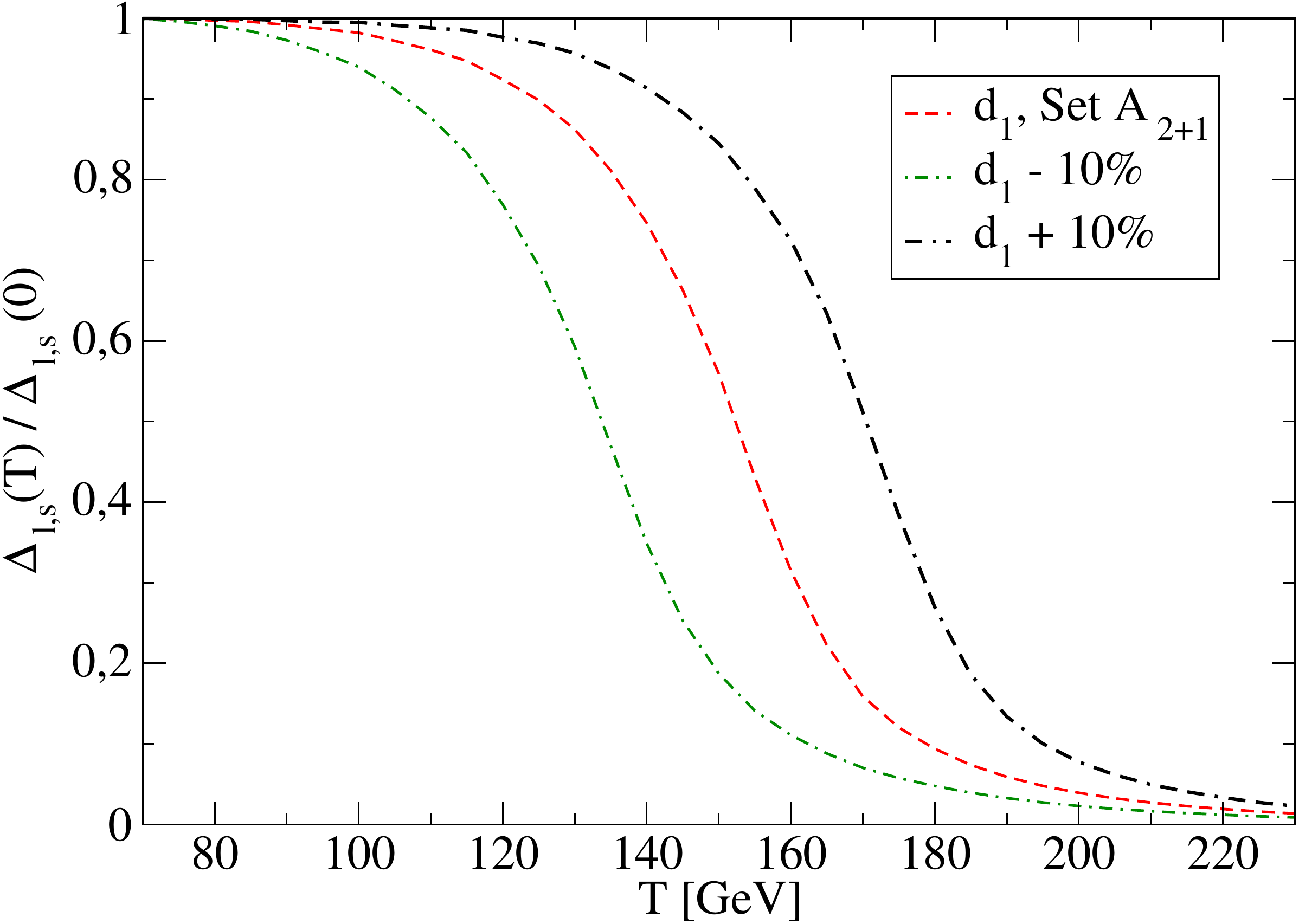}
\caption{\label{fig:varyd1}Influence of $d_1$ on $\Delta_{l,s}$.}
\end{minipage} 
\end{figure}

In Fig.~\ref{fig:mu000} we show the regularized chiral condensate at $\mu$=0 MeV for all sets of interest, compared to lattice data taken from \cite{Borsanyi:2010bp}. 
We see a nice matching of Set A$_{2+1}$ (dashed red line) with the lattice data, not only in the scale ($T_C$; 
controlled by the choice of $d_1$ as shown in Fig.~\ref{fig:varyd1}) but also in the steepness of the regularized condensate, which is a non-trivial result.
Additionally we observe that by merely adding the charm quark in Set A (dash-dotted blue line), we find a shift of the scale to lower temperatures, which is an effect caused by adding the charm quark
to the system without adjusting the physical scale. To investigate this, by looking at
Sets B (solid black line) we find that by self-consistently readjusting the scale, the charm quark has no effect. 

\begin{figure}[h]
\begin{minipage}{20pc}
\includegraphics[width=18pc]{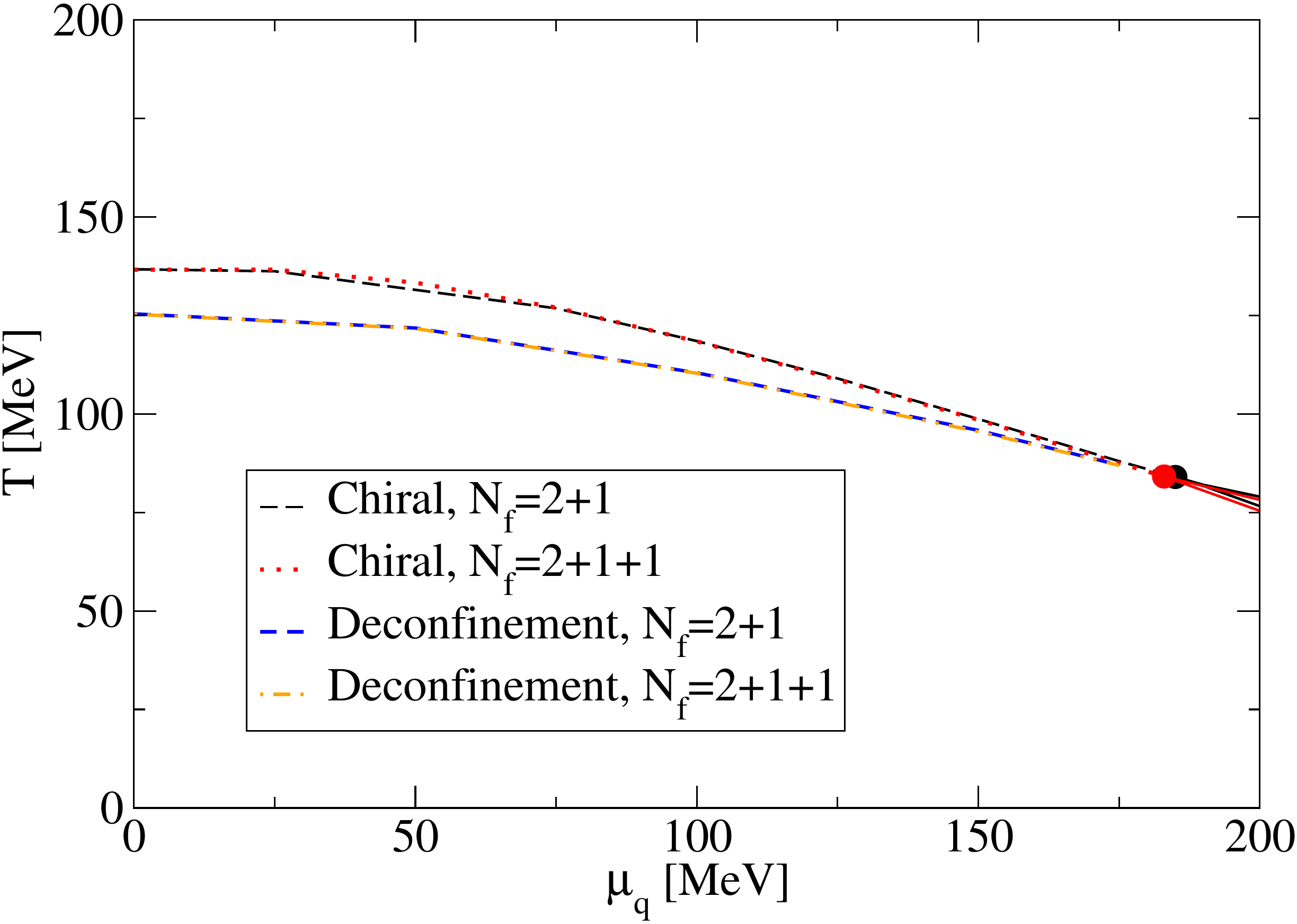}
\caption{\label{fig:pdSetB}Phase diagram Sets B \cite{Fischer:2014ata}.}
\end{minipage}\hspace{0pc}%
\begin{minipage}{20pc}
\includegraphics[width=18pc]{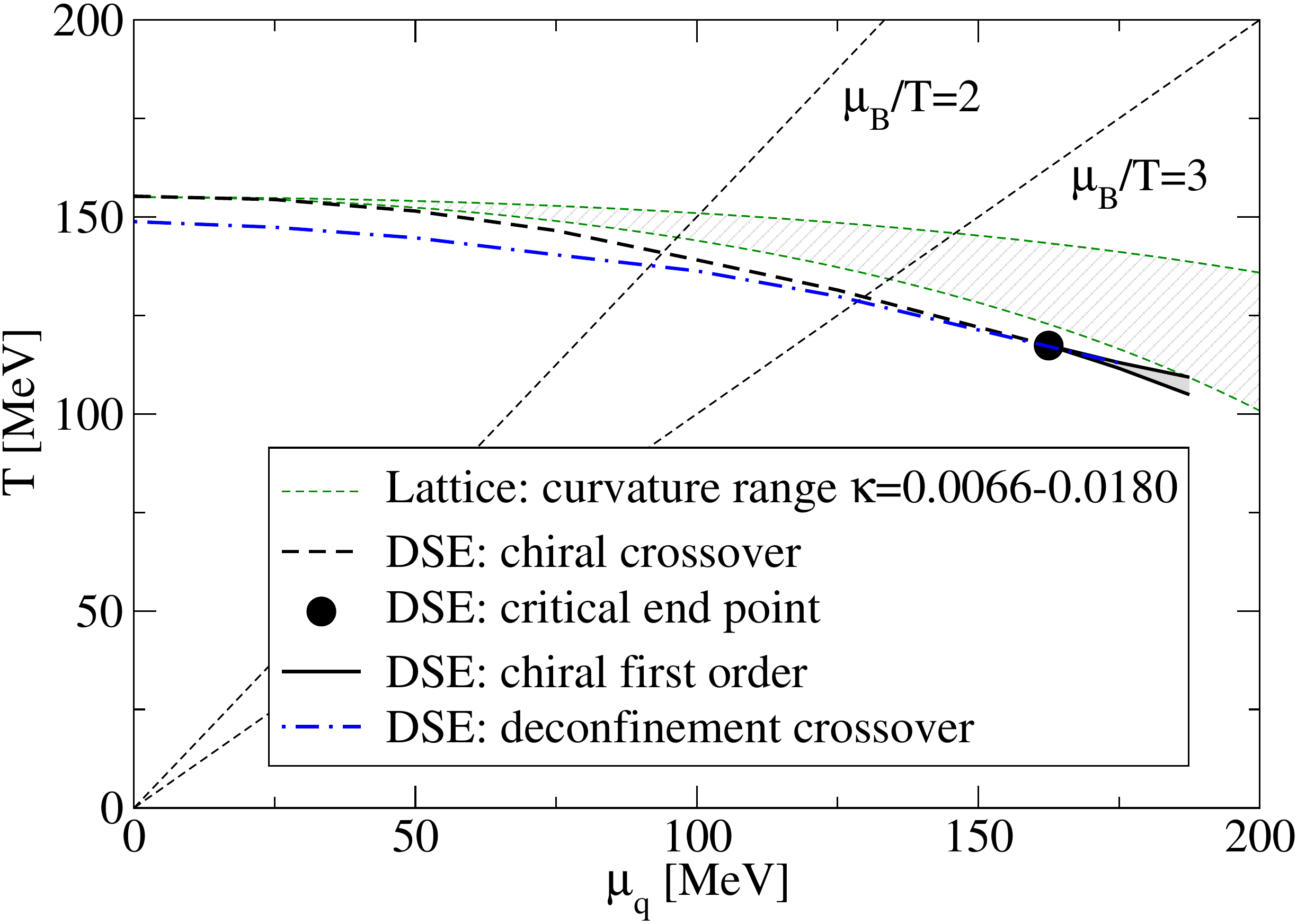}
\caption{\label{fig:pdSetA}Phase diagram Set $A_{2+1}$ \cite{Fischer:2014ata}.}
\end{minipage} 
\end{figure}

Fig.~\ref{fig:pdSetB} shows the phase diagrams for Sets B. It immediately becomes apparent, that the negligible contribution of the charm quark which we observed for $\mu$=0 MeV in Fig.~\ref{fig:mu000}, continues to be small for the whole T-$\mu$-plane for 
both, chiral and deconfinement transition. The deconfinement transition was fixed via the Polyakov loop of the minimum of the background field potential (see \cite{Fischer:2013eca} for more details), where the inflection point is used. On the other hand, 
the critical temperature in the chiral crossover region was defined by the maximum of the chiral susceptibility. This partly explains the gap for small chemical potential between the crossover lines. 

Our results for Fig.~\ref{fig:pdSetB} suggest that there is almost no influence of the charm quark on the  
transition temperatures, provided the internal scales are controlled by external input.
For the QCD phase diagram shown in Fig.~\ref{fig:pdSetA} we therefore return to lattice input for this 
task (Set A$_{2+1}$). In this plot,
the critical temperature for the crossover region is in both cases, chiral and deconfinement transition, defined by the inflection point method. One can see, that the chiral crossover (dashed black line) becomes ever steeper with increasing
chemical potential and turns into a critical end point (CEP) at 
\begin{equation}
(T^c,\mu_q^c)=(115,168) \,\mbox{MeV}. 
\end{equation}
To guide the eye we also show the lines for $\mu_B$/T=2 and $\mu_B$/T=3 and added predictions for the curvature of the chiral transition line from lattice 
extrapolations for $N_f=2+1$ of different groups at imaginary and zero chemical potential into the real chemical potential region \cite{Endrodi:2011gv,Kaczmarek:2011zz,Cea:2014xva}.
The agreement between the lattice extrapolation and our DSE results is quite satisfactory.
We close with the remark that potential effects of baryons onto the location of the critical
endpoint have not yet been taken into account, neither in the lattice extrapolations of
the curvature, nor in our DSE approach. In order to turn this qualitative study into a quantitative
one, these need to be addressed. Thus the close agreement of both approaches, although interesting,
may very well not be the final word. 

\vspace*{3mm}
{\bf Acknowledgements}\\ This work has been supported by the Helmholtz Alliance
HA216/EMMI and by ERC-AdG-290623 as well as the Helmholtz
International Center for FAIR within the LOEWE program of the State of
Hesse.


\end{document}